\documentclass[aps,prl,reprint,showpacs,groupedaddress,letterpaper]{revtex4-1}
\usepackage{amsmath,amssymb,bm}
\usepackage{dcolumn,longtable,graphicx}
\usepackage{gensymb,esvect}
\usepackage[normalem]{ulem}
\usepackage{hyperref}

\def\figsdir{.}

\def\figwidth{3.in}

\newcommand{\pow}{PoW}
\newcommand{\pom}{PoM}
\newcommand{\bigdelim}[3]{\bigl#1#3\bigr#2}

\begin{document}

\title{A network-dependent rewarding system: proof-of-mining}

\author{Joe Lao}
\email[]{CoinMagiCorp@gmail.com}
\affiliation{CoinMagi Corporation, Georgia, Atlanta, USA}

\date{\today}

\begin{abstract}
A 'soft' control of the network activity through varying reward in a proof-of-work (PoW) cryptocurrency is reported. Rewards are the necessity to incent the contributors' activities (i.e., mining) in order to maintain the \pow\ network. Contrary to constant rewarding in a certain period implemented in most of cryptocurrency, such as bitcoin, we propose a network-dependent rewarding model system, primarily including two phases: 1) activities encouraging phase in which higher rewards are issued at higher network activities; and 2) discouraging further increase of activities by reducing rewards. The advantages of this system include 1) fair distribution of rewards among a variety of contributors, and 2) enforcing a limit to the network activity and hence the cost of maintaining the \pow\ network. This mechanism requires network contributors to show their participation in order to earn maximum rewards, i.e., proof-of-mining. 
\end{abstract}
\maketitle

\section{Introduction}
Maintaining the public ledger (i.e., the block chain) that records financial transactions is essential to operate a cryptocurrency system, \cite{cryptocurrency.url} such as bitcoin, \cite{Nakamoto2009} without interruption. Neither under a centralized bank nor under a government, sustaining the block chain (so called mining) is fully driven by volunteer-based activities, through rewarding contributors proportional to their work being done, i.e., proof-of-work (PoW). This process should be accomplished without the intermediation of central authority in order to fulfill decentralization. Work-for-pay means higher rewards are given to one with more work accomplished. This is acceptable in general, however, concerns have been raised regarding the profit-driven hardware competition race, and the global energy consumption; \cite{Dagger.url} in particular the construction of powerful hardwares is changing the decentralized and more open nature of cryptocurrency. \cite{bitcoin.threats.url} There has been several proposed 'hard' approaches reported by using memory-hard algorithm, which increases the difficulty of accomplishing PoW works, thus resistant to specialized hardwares. Expected will be a reduction of network activities and thus incentive to generic hardwares. However, under the present rewarding system with constant rewards over time, it is still tempting for one to build up mining equipment and consequently induces a race in the generic-hardware level, which is also against the decentralization nature. 

In this article, we propose a new block rewarding model system, operating based on the network activity. The activity refers to the mining process, for example in the bitcoin network, which creates new blocks. The concept behind the rewarding system in order to achieve fair distribution (decentralization) and suppressing mining hardware race is achieved by encouraging rewards to higher activities when the network activity on the whole is low, and by discouraging miners through reducing rewards when more powerful equipment comes into the network. It is expected that the block rewards heavily interact with the miners' activities, leading to a large fluctuations in the rewards due in part to the miners leaving and rejoining. To acquire maximum rewards, one has to remain an extended period of mining activities in the network, i.e., to show the proof of mining (\pom). 

\section{A network-dependent rewarding system}

\begin{figure}[!h]
\includegraphics[width=\figwidth]{\figsdir/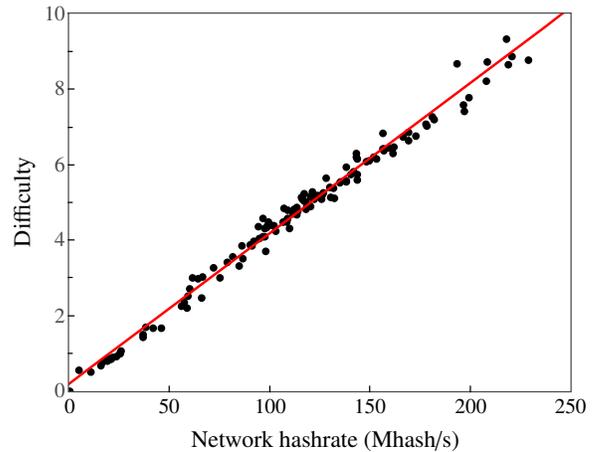}
\caption{\label{fig:xmg.nethash} The network difficulty as a function hashrate obtained from XMG. \cite{Magi.url}
}
\end{figure}

The difficulty to find a block is used as a measure of the network activities. Miners keep finding a target number (hash) acceptable to the network in order to forge the public ledger (block), the efficiency of which is evaluated by the hash rate. Finding new blocks only occurs at pre-designed time slots, which enforces an adjustment of the block difficulty in accordance with varying hash rate in the network. Typical relationship between difficulty and network hashrate extracted from Coin Magi (XMG) \cite{Magi.url} is shown in Fig.~\ref{fig:xmg.nethash}. 

\begin{figure}[!h]
\includegraphics[width=3.3in]{\figsdir/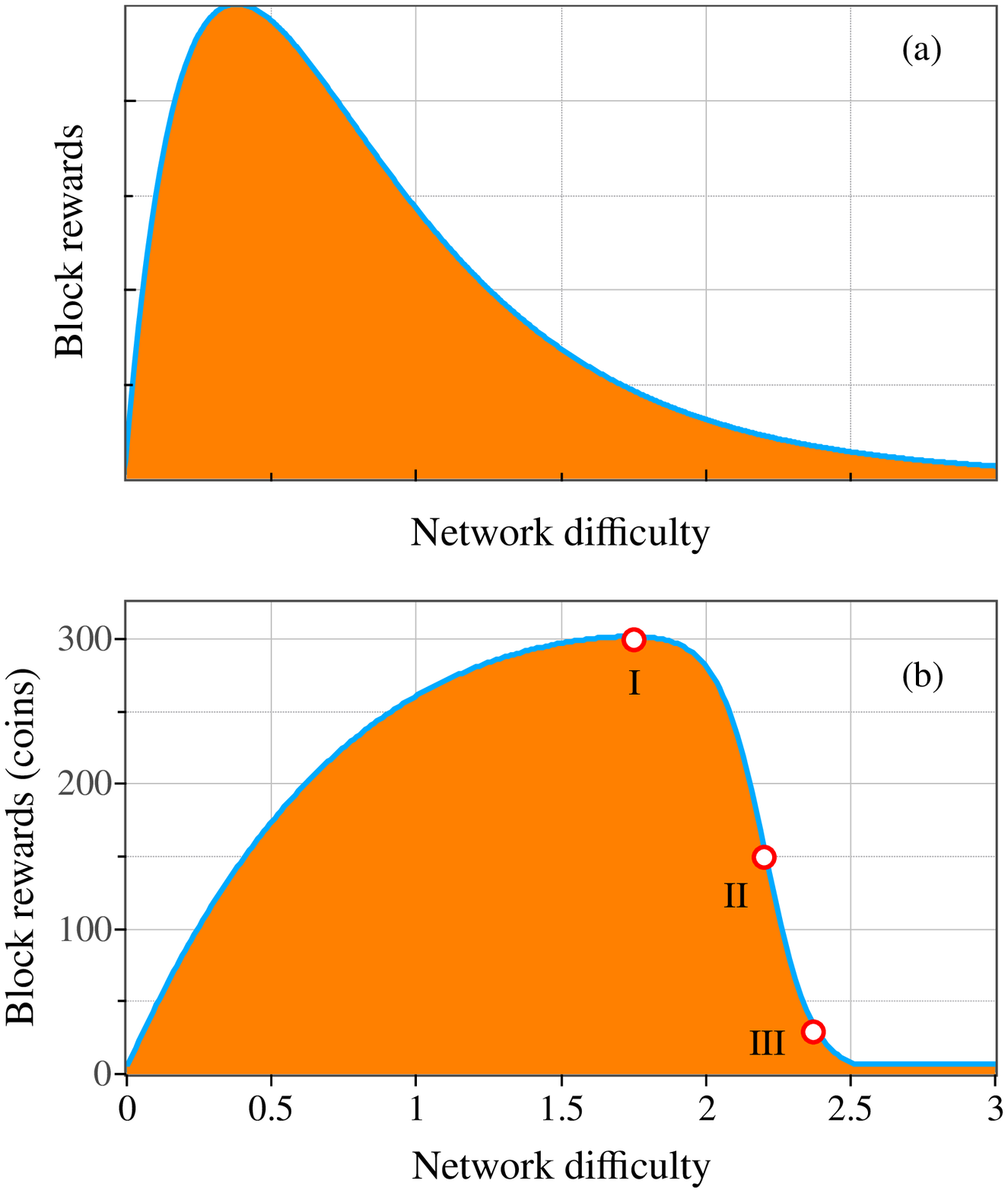}
\caption{\label{fig:rewards} (a) The variation of block rewards with network difficulty in terms of Eq.~\ref{eq:reward}. (b) The actual block rewards used in a cryptocurrency, XMG. \cite{Magi.url}. The block reward at the high-$\mathcal{D}$ side degrades much faster compared to (a), which leads to efficient adjustment of the network activities. For example, block reward is halved when $\mathcal{D}=2.20$ (II) compared to the value at I, and equals to 1/10 of the maximum value at $\mathcal{D}=2.37$, at which the low rewards will be lack of incentive to the miners. }
\end{figure}

The proposed block reward ($\mathcal{R}$) correlates with the difficulty ($\mathcal{D}$) through
\begin{equation}\label{eq:reward}
\mathcal{R}
\sim\sqrt{\bigdelim{[}{]}{\exp(-a\cdot\mathcal{D})-\exp(-b\cdot\mathcal{D})}\cdot\mathcal{D}},~~(a < b)
\end{equation}
where $a$ and $b$ are empirical parameters. $\mathcal{R}$ relies on $\sqrt{\mathcal{D}}$ for small values of $\mathcal{D}$, with the line shape tuned by the exponential functions; at large values of $\mathcal{D}$, $\exp(-a\cdot\mathcal{D})$ dominates. Fig.~\ref{fig:rewards} (a) shows a representative variation of the block rewards with $\mathcal{D}$. 

\section{Discussion}

The rising part of Fig.~\ref{fig:rewards} (a) is of importance to maintenance a fair distribution of rewards into public. Since single person has limited hashing power, we can consider the network hashrate to be proportional to the number of miners. It is apparent that the number of miners could significantly changed over lifetime of mining. Issuing the same rewards despite that the network consists of 1 or 1000 miners, would result in unfairness. Furthermore, the amount of reward is disproportional to the hashing power by noticing such a fact that one who has same equipment could gain high reward at the early stage of mining when only a few people are involved. The situation could be even worse if the mining reward is designed to be decreasing with $\mathcal{D}$; in this case, the initial miners will acquire significant high percentage of rewards compared to the subsequent miners. It is therefore very natural that the block reward should be designed such that it increases with the number of miners. The early miners still have the advantage of incremental rewards over time; this is reasonable to enable rewarding miners' early participation. 

Apparently increasing reward can drive miners towards building more powerful equipment. Such a consequence has to be mitigated, which is achieved by the mining discourage phase at the high-$\mathcal{D}$ range. The relationship between mining costs and returns in consistence with economics should be the guideline to achieve the goal. We assume this is valid throughout this investigation. For example, one who has powerful equipment and hence high cost as well will expect commensurate returns; otherwise, the cost must be reduced by reducing the usage of equipment. The efficiency of discouraging mining activities is fully determined how fast the rewards decline with $\mathcal{D}$. 

It should be noted that self-adjustment of network difficulty in accordance with rewards is affected by both the reward declining rate and the market place of the cryptocurrency in addition to the declining rate. One can build up more powerful equipment to compensate the declining rewards. A much faster declining rate will disable any possibility of acquiring desired rewards by simply building more hashing power. Fig.~\ref{fig:rewards} (b) shows an improved rewarding system as implemented in XMG. The high-$\mathcal{D}$ range is determined by a cut-off function, similar to the Fermi-Dirac function used in the physical science:
\begin{equation}\label{eq:fd}
\mathcal{F}_{FD}^{CO}
=1/\bigdelim{[}{]}{1+\exp\bigdelim{(}{)}{\frac{\mathcal{D}-\mathcal{D}_{CO}}{\alpha\cdot T}}}
\end{equation}
$\mathcal{D}_{CO}$ determines where the decline occurs. $T$ is a quantity of affecting the spread of the declining range. As for XMG, the maximum block reward is available at $\mathcal{D}=1.75$ (I) (or $\sim40$~MHash/s according to Fig.~\ref{fig:xmg.nethash}). The reward is halved when $\mathcal{D}=2.20$ (or $\sim51$~MHash/s)  (II), and equals to 1/10 of the maximum value at $\mathcal{D}=2.37$ (or $\sim55$~MHash/s), at which the low rewards will be lack of incentive to the miners. It has been proven that the ratio of the hashrate of miners higher than 5~Mhash/s over the total network hashrate is decreased from 47.5~\% to 38.4~\% upon the implementation of the new rewarding system, while the network hashrate decreases from 115~MHash/s to 83~MHash/s. These quantities are still in varying with time. It can be concluded that the new block rewarding system has a significant effect on the miners' activities. Continuous increasing of the hash power has been hindered.  

One may expect the market place will affect the miners' behavior significantly. For example, the high-pricing reward leads more miners into the network which immediately move the block reward out of the maximum region. This initiates a self-adjustment of the  rewarding system. Realizing the marginal rewards (e.g., at III of Fig.~\ref{fig:rewards} (b)), some of the miners start ceasing their minings, thus pushing the reward back to the maximum. In this case, it will be relatively less advantageous for miners quickly in and out the mining activity. One will need to show a proof of mining in order to gain sufficient credits to declare the rewards. 

\section{Conclusion}

To summary, we have proposed a new block rewarding system to achieve fair distribution of block rewards among miners and to prevent mining equipment competition as seen in cryptocurrency, such as bitcoin. Under this system, earlier adopters don't have significant advantages over the latter adopters. High hashrate is discouraged through adjusting the rewards: high rewards are issued at low network hashrate. It is also observed that the new system emphasizes the participation of miners instead of their hashing power, allowing for a new mining mechanism, i.e., proof-of-mining. It is anticipated that the new block rewarding system leads to a more decentralized cryptocurrency. 




\end{document}